\documentclass[10pt]{article}
\usepackage[utf8]{inputenc}
\date{24/05/19}
\usepackage{amsmath}
\usepackage{amsfonts}
\usepackage{amssymb}
\usepackage{extsizes}
\usepackage{graphicx}
\usepackage{tikz}
\usepackage{authblk}
\usepackage[left=2cm,right=2cm,top=2cm,bottom=4cm]{geometry}

\date{\today}
\title{\textbf{Spectral triple with real structure on fuzzy sphere}}
\author[1]{Anwesha Chakraborty\thanks{anwesha@bose.res.in }}
\author[1]{Partha Nandi \thanks{parthanandi@bose.res.in}}
\author[2]{Biswajit Chakraborty		\thanks{dhrubashillong@gmail.com}}
\affil[1]{Department of Theoretical Sciences\\
	
	S.N. Bose National Centre for Basic Sciences\\
	
	JD Block, Sector III, Salt Lake, Kolkata 700106, India}
\affil[2]{Department of Physics\\
	
	Ramakrishna Mission Vivekananda Educational and Research Institute\\
	
	PO-Belur Math, Howrah-711202, West Bengal, India}
\begin{document}
	\maketitle
	\begin{abstract}
		Here we have illustrated the construction of a real structure on fuzzy sphere $S^2_*$ in its spin-1/2 representation. Considering the SU(2) covariant Dirac and chirality operator on $S^2_*$ given by Watamura et. al. in [U. C Watamura, S Watamura, \textit{Comm. Math. Phys., \textbf{183} P-365–382 (1997)}; \textit{Comm. Math. Phys.  \textbf{212} P-395-413(2000)}], we have shown that the real structure is consistent with other spectral data for KO dimension-4 fulfilling the zero order condition, where we find it necessary to enlarge the symmetry group from SO(3) to the full orthogonal group O(3). However the first order condition is violated thus paving the way to construct a toy model for an SU(2) gauge theory to capture some features of physics beyond standard model following Connes et.al. [A. H. Chamseddine, A. Connes and W. D. van Suijlekom, \textit{JHEP 1311, \textbf{132} (2013)}].
	\end{abstract}
	\section{Introduction and motivation}
	%	The idea is to construct an SU(2) gauge theory , \textit{a la'} Connes et. al. , by taking the internal space to be the fuzzy sphere $S_*^2$ and then study some contrasting features of this theory by comparing with the well known Kaluza-Klein theory , where the internal space is taken to be the commutative round sphere $S^2$.
	Fuzzy sphere (FS) $S ^2_*$ is a simple example of noncommutative geometry (NCG),  introduced a long time back by J. Madore \cite{mad} which can be thought of as a deformation of the usual commutative 2-sphere $S^2$. Recently, there have been a flurry of activities involving fuzzy sphere in the high energy physics community as this provides yet another non-perturbative technique in quantum field theory \cite{grosse} and which is alternative to the lattice gauge theory, where the latter is based on  finite-discretization of spacetime. The main advantage is that the algebra of FS can carry representations of SU(2) Lie group, which is given by $M_{2n+1}$ matrices for its n-th representation, so it is fruitful to study the geometry in context of both matrix models and NCG.  \\
	Our aim in this paper, is to just construct real  structure on a fuzzy sphere, so that we can have a real and even spectral data in our possession. This in turn, should pave the way to construct a simple toy model having SU(2) gauge symmetry by taking the internal space to be a fuzzy sphere $S^2_*$ in an almost commutative geometrical framework, which we would like to take up as our future work .  Although, there are various proposals of Dirac operators on fuzzy sphere \cite{mad,gros,promod,yadri}, along with the grading operator in some cases, the real structure on fuzzy sphere seems to be the most important missing ingredient. In this backdrop we have considered the SU(2) covariant Dirac operator and grading operator on $S^2_*$ proposed by Watamura et.al. in \cite{wat} and made an attempt to construct the real structure on fuzzy sphere in its lowest spin-1/2 representation. The $S^2_*$ in its 1/2 representation is a very special one in the family of fuzzy spheres. Not only
	that it represents a space of maximal noncommutativity, the algebra $\mathcal{A}_F$ describing this space is given by the
	unital algebra of $2\times 2$ matrices $M_2(\mathbb{C})$ and which is entirely spanned by the 4 generators: $(I, \vec{\sigma})$ i.e.  the identity
	operator and three su(2) generators. Clearly, this feature will not persist for $S^2_*$ associated with higher $n \ge 1$ representations $\rho_n$. As we shall see that this can be leveraged to construct the real structure. But for that we had to enlarge the symmetry group of the fuzzy sphere from SO(3) to O(3)-the total orthogonal group and found that it can be assigned a KO dimension-4.  The Dirac operator, however, violates the first order condition, which  was a necessary condition in the formulation of the standard model of particles in the framework of almost commutative geometry \cite{suij,lizzi}.  In this context, let us point out the literatures \cite{connes3,connes5}, where  Connes et.al. has come up with necessary mathematical formalism  for spectral triples where although the first order condition gets violated, it nevertheless  eventually facilitates the development of a  phenomenologically viable model like Pati-Salam model, which goes beyond standard model. This gives us another motivation to formulate a complete spectral data  for the fuzzy sphere so that one can  use it to construct a toy model, to begin with, to capture some of the aspects of physics beyond the standard model.\\ \\
	The paper is organized as follows: In section-2 we have provided all the ingredients of the spectral triple of the fuzzy sphere, except the real structure i.e. the algebra , the Hilbert space, the Dirac 
	operator and the grading operator for a spin-1/2 fuzzy sphere. As
	said earlier, the latter two structures were borrowed from Watamura et.al. \cite{wat}. Here we found it necessary to introduce a suitable hierarchies of Hilbert spaces.  In section-3, we have derived the eigenspinors of the Dirac operator given in \cite{wat} and also constructed
	the chiral spinors followed by the representation of the algebra and the opposite algebra in various Hilbert spaces
	in section-4. In section-5, we have determined the KO dimension of the theory followed by a detailed derivation
	of the real structure operator and then verified the zero-order condition. The spectral data however, is shown to violate the
	first order condition. We have concluded and discussed the future direction of this work in section-6.
	%In section-5 we have tried to construct super unitary operator defined patch-wise on $S^2_*$, which are helpful to transform the globally SU(2) covariant Dirac and chirality operators to the respective operators defined on a local tangent plane. With help of these previously derived super unitary operators, in section-6, we have shown the 'local' unitary equivalence between the SU(2) covariant chirality operator and $\sigma_3$, which is the chirality operator for a 2 dimensional plane. 
	%Apart from the main text, in Appendix-A.1. we have reviewed the construction of SU(2) principal budle over commutative sphere, which we have emulated for the construction of same for the case of $S^2_*$ and in appendix-A.2, we have shown the consruction of SU(2) perelomov coherent states and the corresponding density matrices for spin-$\frac{1}{2}, 1$ representations. 
	\section{Spectral Triple for $S_*^2$}
	$S_*^2$ algebra is given by 
	\begin{equation}
		[\hat{x}_i,\hat{x}_j] = i\lambda\epsilon_{ijk}\hat{x}_{k};\,\,\,\,\,\,\,\,\,\,i,j,k\in \{1,2,3\} \label{eq1}
	\end{equation}
	Here $\lambda$ is the noncommutative parameter having the dimension of length and the Casimir operator $r_n^2=\rho_n(\hat{\vec{x}}^2)=\rho_n(\hat{x}_i).\rho_n(\hat{x}_i)$ allows only a discrete set of values given by $\lambda^2n(n+1)$, where $n\in \{\frac{1}{2},1,\frac{3}{2},...\}$ is the SU(2) representation index. We shall work here with $n=\frac{1}{2}$ representation only and reserve $\hat{x}_i$ to denote $n=\frac{1}{2}$ representation only: $\rho_{\frac{1}{2}}(\hat{\vec{x}})=\hat{\vec{x}}$ and higher order $(n \ge 1,\frac{3}{2},...)$ representation as $\rho_n(\hat{x})$ explicitly, when need arises. This set of infinite number of fuzzy spheres, indexed by $n$, can be thought foliating noncommutative $R_*^3$ where all of these fuzzy spheres can be associated with their respective radii $r_n=\lambda\sqrt{n(n+1)}$.\\
	Now with $n=\frac{1}{2}$, a representation of the algebra (\ref{eq1})  is furnished by the 2D Hilbert space \begin{equation}
		\mathcal{H}_c=span_{\mathbb{C}}\Big\{\Big|\frac{1}{2}\Big\rangle ,\Big|-\frac{1}{2}\Big\rangle \Big\}\label{e61}
	\end{equation}
	where the actions of $\hat{x}_i$s on $\mathcal{H}_c$ can be written in terms of ladder operators $\hat{x}_{\pm}$ and $\hat{x}_3$ as,
	\begin{align}
		&\hat{x}_+ \Big|\frac{1}{2}\Big\rangle = \hat{x}_- \Big|-\frac{1}{2}\Big\rangle = 0;\,\,\,\,\,\,\,\,
		\hat{x}_+\Big|-\frac{1}{2}\Big\rangle =\lambda \Big|\frac{1}{2}\Big\rangle ; \hat{x}_- \Big|\frac{1}{2}\Big\rangle = \lambda \Big|-\frac{1}{2}\Big\rangle \nonumber\\ 
		&\hat{x}_3 \Big|\pm \frac{1}{2}\Big\rangle = \pm \frac{\lambda}{2} \Big|\pm\frac{1}{2}\Big\rangle;\,\,\,\,\,\,\,\hat{x}_{\pm}=\hat{x}_1\pm i\hat{x}_2 \label{A1}
	\end{align}
	The orthonormality and completeness condition for $\mathcal{H}_c$ are given as,
	\begin{equation}
		\langle m|n\rangle =\delta_{mn};\,\,\,\,\,\,\,\, \Big|\frac{1}{2}\Big\rangle\Big\langle \frac{1}{2}\Big| + \Big|-\frac{1}{2}\Big\rangle\Big\langle - \frac{1}{2}\Big| := \textbf{I}_{\mathcal{H}_c}\in \mathcal{H}_q\,\,\,\,;\,\,\,\,\, m,n\in \{\frac{1}{2},-\frac{1}{2}\}  \label{A2}
	\end{equation}
	where $\mathcal{H}_q$ is  defined below in (\ref{eq2}).\\
	For the algebra of the triplet, we take $\mathcal{A}_F=\mathcal{H}_q$- the space of Hilbert Schmidt (HS) operators acting on $\mathcal{H}_c$:
	\begin{equation}
		\mathcal{A}_F:=\mathcal{H}_q=\mathcal{H}_c\otimes \tilde{\mathcal{H}_c}=span_{\mathbb{C}} \Big\{\Big|\frac{1}{2}\Big\rangle\Big\langle \frac{1}{2}\Big|, \quad\Big|\frac{1}{2}\Big\rangle\Big\langle -\frac{1}{2}\Big|,\quad\Big|-\frac{1}{2}\Big\rangle\Big\langle \frac{1}{2}\Big|,\quad\Big|-\frac{1}{2}\Big\rangle\Big\langle -\frac{1}{2}\Big|\Big\}\label{eq2}
	\end{equation}
	where $\tilde{\mathcal{H}_c}$ is the dual of $\mathcal{H}_c$. These four basis elements of the form $|m\rangle \langle n| := |m,n) $ also satisfies orthonormality and completeness as (\ref{A2}),
	\begin{equation}
		(m',n'|m,n)=\delta_{m,m'}\delta_{n,n'};\,\,\,\, \sum_{m,n} |m,n)(m,n| =\textbf{I}_{\mathcal{H}_q}  \in \mathcal{H}_q \otimes  \tilde{\mathcal{H}_q}\label{A3}
	\end{equation} 
	where 
	\begin{equation}
		(m',n'|=|m',n')^*=(|m'\rangle \langle n'|)^*= |n'\rangle \langle m'| \label{A4}
	\end{equation}
	Here  the inner product between a pair of HS operators $|\psi), |\phi)\in \mathcal{H}_q$ is defined as ,
	\begin{equation}
		(\phi|\psi):=tr_{\mathcal{H}_c}(\phi^*\psi)= \sum_{i=-\frac{1}{2}}^{+\frac{1}{2}}
		\langle i|\phi^*\psi|i\rangle ; \,\,\,\, |i\rangle \in \mathcal{H}_c \label{eq5}
	\end{equation}
	with *- operation being the involution operation of the algebra $\mathcal{A}_F=\mathcal{H}_q$ and corresponds to the simple hermitian conjugation in this case fulfilling the following properties
	\begin{equation}
		(a^*)^* = a ;\,\,\,(ab)^*=b^*a^*\,\,\,\,\forall a, b\in \mathcal{A}_F \label{N44}
	\end{equation}
	One can expect to identify the inner automorphism symmetry of $\mathcal{A}_F$ with the gauge symmetry  of $\mathcal{A}_F$.
	Regarding the Hilbert space of the spectral triple (which should be a bi-module of the algebra) , we take it as $\mathbb{C}^2 \otimes \mathcal{H}_q$ i.e. 
	\begin{equation}
		\mathcal{H}_F:=\mathbb{C}^2 \otimes \mathcal{H}_q=span_{\mathbb{C}} \{|\phi_1)),|\phi_2)),|\phi_3)),|\phi_4)),|\phi_5)),|\phi_6)),|\phi_7)),|\phi_8))\} \label{eq3}
	\end{equation}
	where
	\begin{align}
		&|\phi_1))=\begin{pmatrix}
			\Big|\frac{1}{2}\Big\rangle\Big\langle \frac{1}{2}\Big|\\0
		\end{pmatrix},\,\,\,\,
		|\phi_2))=
		\begin{pmatrix}
			\Big|\frac{1}{2}\Big\rangle\Big\langle -\frac{1}{2}\Big|\\0
		\end{pmatrix},\,\,\,\,
		|\phi_3))=\begin{pmatrix}
			\Big|-\frac{1}{2}\Big\rangle\Big\langle \frac{1}{2}\Big|\\0
		\end{pmatrix},\,\,\,\,
		|\phi_4))=\begin{pmatrix}
			\Big|-\frac{1}{2}\Big\rangle\Big\langle -\frac{1}{2}\Big|\\0
		\end{pmatrix} \nonumber \\ 
		&|\phi_5))=\begin{pmatrix}
			0\\\Big|\frac{1}{2}\Big\rangle\Big\langle \frac{1}{2}\Big|
		\end{pmatrix},\,\,\,\,
		|\phi_6))=
		\begin{pmatrix}
			0\\\Big|\frac{1}{2}\Big\rangle\Big\langle -\frac{1}{2}\Big|
		\end{pmatrix},\,\,\,\,\,|\phi_7))=\begin{pmatrix}
			0\\\Big|-\frac{1}{2}\Big\rangle\Big\langle \frac{1}{2}\Big|
		\end{pmatrix},\,\,\,\,|\phi_8))=
		\begin{pmatrix}
			0\\\Big|-\frac{1}{2}\Big\rangle\Big\langle -\frac{1}{2}\Big|
		\end{pmatrix} \label{N2}
	\end{align}
	where the $\{|\phi_{\mu})) ; \mu= 1,2,...,8\}$ furnish a complete and orthonormal basis for $\mathbb{C}^2\otimes\mathcal{H}_q$:
	\begin{equation}
		((\phi_{\mu}|\phi_{\mu^{\prime}})) =\delta_{\mu\mu^{\prime}};\,\,\,\,\,\,\,\, \sum_{\mu=1}^8 |\phi_{\mu}))\,(( \phi_{\mu}|=\textbf{I}_{\mathcal{H}_F} \in {\mathcal{H}_F \otimes \tilde{\mathcal{H}}_F}\label{eq4}
	\end{equation}
	Here the inner-product between a pair of elements $|\psi_1)) , |\psi_2)) \in \mathcal{H}_F=\mathbb{C}^2\otimes\mathcal{H}_q$ has been defined as,
	\begin{equation}
		((\psi_1 |\psi_2))= tr_{\mathcal{H}_c\otimes\mathbb{C}^2}(\psi_1^{\dagger}\psi_2)\label{eq6}
	\end{equation} 	  
	where we have reserved the dagger ($\dagger$) symbol for its use in $\mathcal{H}_F$:
	\begin{equation}
		((\psi |:= |\psi))^{\dagger}=(\,\,(\xi_1|\,\,\,\,\,\,\,\,\, (\xi_2|\,\,) \,\,\,\textrm{for} \,\,\,|\psi)) :=\begin{pmatrix}
			|\xi_1)\\ |\xi_2)
		\end{pmatrix};\,\,\,\, |\xi_1),|\xi_2)\in \mathcal{H}_q,\,\,\textrm{and}\,\,\, (\xi_1|, (\xi_2| \in \tilde{\mathcal{H}}_q \label{eq7}
	\end{equation}
	We shall refer to the basis $|\phi_{\lambda})) \,\,\,(\lambda=1,2,....,8)$ as the canonical basis. Note that here we have a hierarchies of Hilbert spaces $\mathcal{H}_c, \mathcal{H}_q= \mathcal{H}_c\otimes \tilde{\mathcal{H}}_c $ and $\mathcal{H}_F=\mathbb{C}^2\otimes\mathcal{H}_q$ and the respective elements are denoted by $|. \rangle, |. ), |. )) $ . Further the Hilbert space $\mathcal{H}_F$ can be regarded as the module (in-fact a bi-module in our case as the right action can also be defined) of the algebra $\mathcal{A}_F$, through the diagonal representation $\pi(a)=diag(a,a)\,\,\forall\,a\in\mathcal{A}_F$.\\
	Finally , the Dirac operator and the chirality operator are given by \cite{wat} 
	\begin{equation}
		D_F= \frac{i}{\lambda r_n} \gamma_F\epsilon_{ijk}\sigma_i\hat{x_j}^R\hat{x}_k;\,\,\,\,\,\, \gamma_F=\frac{1}{\mathcal{N}}(\vec{\sigma}.\hat{\vec{x}}^R-\frac{\lambda}{2}) \label{eq8}
	\end{equation}	
	Here the superscript R represents the right action of the operator $\hat{x}_j$. 
	In absence of any superscript ,the action of $\hat{x}_i$ is taken to be from the left- by default; the superscript L is suppressed in this case. \\Although, our analysis will be restricted here for $n=\frac{1}{2}$ representation, we would like to mention here that for general case, $r_n$ is the radius of $n$-th fuzzy sphere as mentioned below (\ref{eq1}) and $\mathcal{N}=\lambda (n+\frac{1}{2})$  serves as a normalisation constant for $\gamma_F$ in (\ref{eq8}) , which is mandated to satisfy $\gamma_F^2=1$. For n= $\frac{1}{2}$ case they simply become,
	\begin{equation}
		r_{\frac{1}{2}}^2= \frac{3\lambda}{4}\,\,\,\textrm{and} \,\,\,\, \mathcal{N}=\lambda\label{N45}
	\end{equation} 
	This algebra $\mathcal{A}_F$ (\ref{eq2}), $\mathcal{H}_F$ (\ref{eq3}) and $D_F$ (\ref{eq8}) furnishes the three primary ingredients for our spectral triple of the fuzzy sphere . In order to formulate a gauge theory in the framework of an almost commutative manifold, we can therefore first compose this spectral triple with these three ingredients  (\ref{eq2},\ref{eq3},\ref{eq8}) with those of a compact Euclidean manifold $M_4$ to get the composite spectral triple as
	\begin{equation}
		\mathcal{A}=C^{\infty}(M_4,\mathcal{A}_F);\,\,\,\, \mathcal{H}=L^2(M_4,\mathcal{H}_F)=L^2(M_4)\otimes \mathcal{H}_F;\,\,\,\,D=D_M\otimes \textbf{I}+\gamma_5\otimes D_F \label{eq11}
	\end{equation} 
	In addition to these data , we need also to have the total grading operator $\gamma$ and the real structure $\mathcal{J}$. The former is given by
	\begin{equation}
		\gamma= \gamma_5\otimes\gamma_F\label{eq12}
	\end{equation}
	where $\gamma_F$ is given in (\ref{eq8}). Regrading the latter, we first need to have the real structure $\mathcal{J}_F$ for $S_*^2$ first, so that the composite real structure can be written as,
	\begin{equation}
		\mathcal{J}=\mathcal{J}_M\otimes\mathcal{J}_F \label{eq13}
	\end{equation} 
	where $\mathcal{J}_M$ is the real structure for $M_4$ and its action is simply given by a complex conjugation.
	\section{Determination of eigenspinors of $\mathcal{D}_F$ and the chiral basis}
	In order to obtain $\mathcal{J}_F$, we first need to obtain the eigen-spinors of the Dirac operator $D_F$ (\ref{eq8}) . Now since the Hilbert space $\mathbb{C}^2 \otimes \mathcal{H}_q$ is 8-dimensional, we expect to have 8 linearly independent eigen-spinors, which can also provide an orthonormal complete set of basis for $\mathbb{C}^2 \otimes \mathcal{H}_q$ and will serve as an alternative to the canonical basis $|\phi_{\lambda}))\,\,\,\, (\lambda \in \{1,2,....,8\})$. To this end ,one can take as an ansatz
	\begin{equation}
		|\psi)) = \begin{pmatrix}
			\sum_{m,n}A_{m,n}|m\rangle\langle n|\\
			\sum_{p,q}B_{p,q}|p\rangle \langle q| 
		\end{pmatrix}\,\,\in \mathcal{H}_q\otimes\mathbb{C}^2;\,\,\,\,n,m,p,q\in \{\frac{1}{2},-\frac{1}{2}\} \label{eq14}
	\end{equation}
	satisfying the eigenvalue equation
	\begin{equation}
		D_F |\psi)) = m|\psi)) \label{eq15}
	\end{equation}
	and solve it. \\
	Further since $[D_F,\hat{\vec{J}}]=0$, where $\hat{\vec{J}}=\{J_1,J_2,J_3\}$ are the 3-components of the total angular momentum  \begin{equation}
		\hat{J}_i=\hat{L}_i+\frac{\sigma_i}{2};\,\,\,L_i=\frac{1}{\lambda}(\hat{x}_i-\hat{x}^R_i) \label{eq16}
	\end{equation}
	where $\vec{L}$ can be regarded as the orbital angular momentum and like $\vec{J}$ this too is dimensionless. It is clear at this stage that these eigen spinors can be lebelled by their respective 'mass'  $m$ (\ref{eq15}) , $\vec{J}^2,J_3$ eigenvalues as $|m;j,j_3))$ satisfying,
	\begin{equation}
		D_F|m;j,j_3))= m |m;j,j_3));\,\,\,\,\,\hat{J}_3|m;j,j_3))= j_3 |m;j,j_3));\,\,\,\,\,\hat{\vec{J}}^2|m;j,j_3))= j(j+1)|m;j,j_3))\label{eq17}
	\end{equation}
	A long but straight forward calculation yields the following structures for the normalised eigen spinors,
	\begin{align}
		|\psi_1)) &= \frac{1}{3-\sqrt{3}}\begin{pmatrix}
			\Big|\frac{1}{2}\Big\rangle\Big\langle \frac{1}{2}\Big| +(\sqrt{3}-1) \Big|-\frac{1}{2}\Big\rangle\Big\langle -\frac{1}{2}\Big|\\
			(2-\sqrt{3})\Big|\frac{1}{2}\Big\rangle\Big\langle -\frac{1}{2}\Big|
		\end{pmatrix}=|1;\frac{1}{2},\frac{1}{2}\rangle \nonumber \\\nonumber\\
		|\psi_2)) &=\frac{1}{3-\sqrt{3}} \begin{pmatrix}
			(2-\sqrt{3})\Big|-\frac{1}{2}\Big\rangle\Big\langle \frac{1}{2}\Big|\\
			(\sqrt{3}-1)\Big|\frac{1}{2}\Big\rangle\Big\langle \frac{1}{2}\Big|+\Big|-\frac{1}{2}\Big\rangle\Big\langle -\frac{1}{2}\Big|
		\end{pmatrix}=|1;\frac{1}{2},-\frac{1}{2}\rangle \nonumber\\\nonumber\\
		|\psi_3)) &= \frac{1}{3-\sqrt{3}}\begin{pmatrix}
			(1-\sqrt{3})\Big|-\frac{1}{2}\Big\rangle\Big\langle -\frac{1}{2}\Big|+(2-\sqrt{3})\Big|\frac{1}{2}\Big\rangle\Big\langle \frac{1}{2}\Big|\\
			\Big|\frac{1}{2}\Big\rangle\Big\langle -\frac{1}{2}\Big|
		\end{pmatrix}=|-1;\frac{1}{2},\frac{1}{2}\rangle \nonumber\\\nonumber\\
		|\psi_4)) &=\frac{1}{3-\sqrt{3}} \begin{pmatrix}
			\Big|-\frac{1}{2}\Big\rangle\Big\langle \frac{1}{2}\Big|\\
			(2-\sqrt{3})\Big|-\frac{1}{2}\Big\rangle\Big\langle -\frac{1}{2}\Big|+(1-\sqrt{3})\Big|\frac{1}{2}\Big\rangle\Big\langle \frac{1}{2}\Big|
		\end{pmatrix}= |-1;\frac{1}{2},-\frac{1}{2}\rangle\nonumber\\\nonumber\\
		| \psi_5)) &= \begin{pmatrix}
			\Big|\frac{1}{2}\Big\rangle\Big\langle -\frac{1}{2}\Big|\\0  
		\end{pmatrix}=|0;\frac{3}{2}\,\,,\frac{3}{2}\rangle ,\hspace{100pt}|\psi_6)) = \begin{pmatrix}
			0\\ \Big|-\frac{1}{2}\Big\rangle\Big\langle \frac{1}{2}\Big| \end{pmatrix}=|0;\frac{3}{2},-\frac{3}{2}\rangle\nonumber\\ \nonumber\\
		|\psi_7)) &= \frac{1}{\sqrt{3}}\begin{pmatrix}
			-\Big|\frac{1}{2}\Big\rangle\Big\langle \frac{1}{2}\Big|+\Big|-\frac{1}{2}\Big\rangle\Big\langle -\frac{1}{2}\Big|\\\Big|\frac{1}{2}\Big\rangle\Big\langle -\frac{1}{2}\Big| \end{pmatrix}=|0;\frac{3}{2},\frac{1}{2}\rangle,\,\,\,\,\,\,\,\,\,\,\,\,\,\,\,\, |\psi_8 )) = \frac{1}{\sqrt{3}}\begin{pmatrix}
			\Big|-\frac{1}{2}\Big\rangle\Big\langle \frac{1}{2}\Big|\\ \Big|\frac{1}{2}\Big\rangle\Big\langle \frac{1}{2}\Big|-\Big|-\frac{1}{2}\Big\rangle\Big\langle -\frac{1}{2}\Big|
		\end{pmatrix}=|0;\frac{3}{2}\,\,,-\frac{1}{2}\rangle \label{B1}
	\end{align}
	As mentioned above, these $|\psi_{\lambda}))$'s $(\lambda \in \{1;2,....,8\})$ too satisfy the orthonormality  and completeness relations like $|\phi_{\lambda}))$'s i.e. (\ref{eq4}) will hold even if we replace $|\phi_{\lambda}))$ by $|\psi_{\lambda}))$.\\ \\
	It should be noted at this stage that both $m=\pm 1$ are doubly degenerate and constitute spin - $\frac{1}{2}$ doublets in the sense that these pair of doublets ($\psi_1$, $\psi_2$) and ($\psi_3$,$\psi_4$) satisfy
	\begin{align}
		&\hat{J}_+|\psi_1)) = \hat{J}_- |\psi_2 )) =0 ;\,\,\,\, \hat{J}_+ |\psi_2)) = |\psi_1)) ;\,\,\,\,\, \hat{J}_- |\psi_1)) =|\psi_2))  \nonumber \\
		&\hat{J}_+ |\psi_3))  =\hat{J}_- |\psi_4)) = 0 ;\,\,\,\,\,\hat{J}_+|\psi_4))  =|\psi_3)) ;\,\,\,\,\,\, \hat{J}_- |\psi_3)) =|\psi_4));\,\,\,\,\, \hat{J}_{\pm}= \hat{J}_1\pm i\hat{J}_2 \label{N1}
	\end{align}
	Like-wise, the mass-less spinors (m=0) have degeneracy 4 and constitutes an angular momentum $j=\frac{3}{2}$ quadruplet. So here too the highest and lowest weight states $|\psi_5))$ and $|\psi_6))$ are annihilated by $\hat{J}_+$ and $\hat{J}_-$ respectively and all the four states here can be connected by multi-fold actions of these ladder  operator $\hat{J}_{\pm}$. \\
	Furthermore ,  the chirality operator $\gamma_F$ can be shown to connect massive spinors across the family as,
	\begin{equation}
		\psi_3 = \gamma_F\psi_1;\,\,\,\,\, \psi_4=\gamma_F\psi_2\label{eq18}
	\end{equation}
	so that the condition $\{\gamma_F,D_F\}=0$ is trivially satisfied. 
	This enables us to introduce the so-called the chiral basis as\\ \\
	\begin{align}
		\chi_1^+&:= \frac{1}{\sqrt{2}}(\psi_1+\psi_3)=\frac{1}{\sqrt{2}}(1+\gamma_F)\psi_1=\frac{1}{\sqrt{2}} \begin{pmatrix}
			\Big|\frac{1}{2}\Big\rangle\Big\langle \frac{1}{2}\Big| \\
			\Big|\frac{1}{2}\Big\rangle\Big\langle -\frac{1}{2}\Big|
		\end{pmatrix}=|1;\frac{1}{2},\frac{1}{2})) \nonumber\\\nonumber\\
		\chi^-_1 & :=\frac{1}{\sqrt{2}}(\psi_1-\psi_3)=\frac{1}{\sqrt{2}}(1-\gamma_F)\psi_1=\frac{1}{\sqrt{6}}\begin{pmatrix}
			\Big|\frac{1}{2}\Big\rangle\Big\langle \frac{1}{2}\Big|+2\Big|-\frac{1}{2}\Big\rangle\Big\langle -\frac{1}{2}\Big|\\
			-\Big|\frac{1}{2}\Big\rangle\Big\langle -\frac{1}{2}\Big| \end{pmatrix}=|-1;\frac{1}{2},\frac{1}{2})) \nonumber\\\nonumber\\
		\chi^+_2 &:=\frac{1}{\sqrt{2}}(\psi_2+\psi_4)=\frac{1}{\sqrt{2}}(1+\gamma_F)\psi_2=\frac{1}{\sqrt{2}}\begin{pmatrix}
			\Big|-\frac{1}{2}\Big\rangle\Big\langle \frac{1}{2}\Big|\\\Big|-\frac{1}{2}\Big\rangle\Big\langle -\frac{1}{2}\Big|
		\end{pmatrix}=|1;\frac{1}{2},-\frac{1}{2})) \nonumber\\\nonumber\\
		\chi^-_2 &:=\frac{1}{\sqrt{2}}(\psi_2-\psi_4)= \frac{1}{\sqrt{2}}(1-\gamma_F)\psi_2 =\frac{1}{\sqrt{6}}\begin{pmatrix}
			-\Big|-\frac{1}{2}\Big\rangle\Big\langle \frac{1}{2}\Big|\\\Big|-\frac{1}{2}\Big\rangle\Big\langle -\frac{1}{2}\Big|+2\Big|\frac{1}{2}\Big\rangle\Big\langle \frac{1}{2}\Big|
		\end{pmatrix}=|-1;\frac{1}{2},-\frac{1}{2}))\label{eq19}
	\end{align}
	Note that here, on the other hand, we have used the notation $|\gamma;j,j_3))$, with $\gamma$ being the eigen value of the chirality  operator $\gamma_F$ to lebel the states : $\gamma_F| \gamma; j,j_3))= \gamma|\gamma; j,j_3))$.\\
	Finally all the zero modes $|\psi_5)),..., |\psi_8)) $
	have negative chirality: $\gamma= -1$. We can therefore write them as,
	\begin{align}
		|\psi_5)) &= \begin{pmatrix}
			\Big|\frac{1}{2}\Big\rangle\Big\langle -\frac{1}{2}\Big|\\0  
		\end{pmatrix}= |-1,\frac{3}{2},\frac{3}{2}))\nonumber\\\nonumber\\
		|\psi_6)) &= \begin{pmatrix}
			0\\ \Big|-\frac{1}{2}\Big\rangle\Big\langle \frac{1}{2}\Big| \end{pmatrix}= |-1,\frac{3}{2},-\frac{3}{2}))\nonumber\\\nonumber\\
		|\psi_7)) &= \frac{1}{\sqrt{3}}\begin{pmatrix}
			-\Big|\frac{1}{2}\Big\rangle\Big\langle \frac{1}{2}\Big|+\Big|-\frac{1}{2}\Big\rangle\Big\langle -\frac{1}{2}\Big|\\\Big|\frac{1}{2}\Big\rangle\Big\langle -\frac{1}{2}\Big| \end{pmatrix}= |-1,\frac{3}{2},\frac{1}{2}))\nonumber\\\nonumber\\
		|\psi_8)) &= \frac{1}{\sqrt{3}}\begin{pmatrix}
			\Big|-\frac{1}{2}\Big\rangle\Big\langle \frac{1}{2}\Big|\\ \Big|\frac{1}{2}\Big\rangle\Big\langle \frac{1}{2}\Big|-\Big|-\frac{1}{2}\Big\rangle\Big\langle -\frac{1}{2}\Big|
		\end{pmatrix}= |-1,\frac{3}{2},-\frac{1}{2})) \label{eq20}\end{align}
	where we have again used the notation $|\gamma;j,j_3))$ with $\gamma=-1$ for all massless modes.\\  \\
	Let us now pause for a while to point out the physical origin for the assignments of different $j_3$ eigen values to the Dirac and
	chiral spinors. It may be recalled at this stage that, the $(2n+1)$-dimensional Hilbert space $\mathcal{H}_c$ for a $n$ -th fuzzy sphere can be identified with a finite
	dimensional subspace of an infinite dimensional Fock space generated by the ladder operators $(\hat{a}_1, \hat{a}^{\dagger}_1)$ and $(\hat{a}_2 ,\hat{a}_2^{\dagger})$ fulfilling $[\hat{a}_{\alpha},\hat{a}_{\beta}^{\dagger}]=\delta_{\alpha\beta}$, of 2D decoupled harmonic oscillator system which is the Hilbert space given by, 
	\begin{equation}
		\mathcal{F}:=Span_{\mathbb{C}}\{|n_1\rangle \otimes |n_2\rangle\equiv |n_1,n_2\rangle = \frac{(\hat{a}_1\dagger)^{n_1}}{\sqrt{n_1!}}\frac{(\hat{a}_2\dagger)^{n_2}}{\sqrt{n_2!}}|0,0\rangle \,\,\,\forall\,\,\,n_1,n_2 \in \mathbb{Z}\}=\oplus_{n \in \{0,\frac{1}{2},1,\frac{3}{2},...\}} \mathcal{F}_n\label{R4}
	\end{equation}
	Here $\mathcal{F}_n=Span\Big\{|n_1,n_2 \rangle\,\Big| \frac{n_1+n_2}{2}=n \Big\}$ for a fixed value of $n$, is the subspace of $\mathcal{F}$, which provides the representation of $n^{th}$ fuzzy sphere and the states belonging to $\mathcal{F}_n$ can be denoted, alternatively as $|n,n_3\rangle$ for $n_3=\frac{n_2-n_1}{2}$. Note that, the Hilbert space $\mathcal{F}$,  splits into direct sum of such subspaces $\mathcal{F}_n$, which  furnishing by themselves the irreducible representation of su(2) Lie algebra for a fixed value of $n$. Now the su(2) Lie algebra generators in that particular representation, fulfilling (\ref{eq1}), can be given by \textquotedblleft{Jordan} Schwinger' map \cite{biswa}, 
	\begin{equation}
		\rho_n(\hat{\vec{x}})=\frac{\lambda}{2}\hat{\xi}^{\dagger}\vec{\sigma}\hat{\xi}\,\,;\qquad\quad \hat{\xi}= \begin{pmatrix}
			\hat{a}_1\\ \hat{a}_2
		\end{pmatrix}\label{e14}
	\end{equation}
	It can be shown that, under the action of the ladder operators $\rho_n(\hat{x}_{\pm})$ made out of (\ref{e14}), $\mathcal{F}_n$ forms an invariant sub-space. The  Casimir operator corresponding to $n^{th}$ representation, satisfy the relation
	\begin{equation}
		\rho_n(\hat{\vec{x}}^2)=\rho_n(\hat{x}_i).\rho_n(\hat{x}_i)=\lambda^2\,n\Big(n+1\Big)
	\end{equation}
	and can be assigned as the value of the square of the radius $r_n$ for the associated $S^2_*$ (see below (\ref{eq1})). \\
	However note that, in our construction, we are working with the Hilbert space $\mathcal{H}_F=\mathbb{C}^2\otimes\mathcal{H}_q$ (\ref{eq3}), where $\mathcal{H}_q= \mathcal{H}_c \otimes\tilde{\mathcal{H}}_c$. So by using Clebsch-Gordon decomposition of SU(2) representation $\mathcal{H}_q$ will  split into direct sum of a 3 (triplet  $\mathcal{F}_1$) and 1 (singlet $\mathcal{F}_0$) -dimensional representation of su(2): $2 \otimes 2= 3 \oplus 1$ as,
	\begin{align}
		&\textrm{Singlet}\qquad\rightarrow\qquad\,\,\,\,\, |\xi):= \frac{1}{\sqrt{2}}\Big[\Big|\frac{1}{2}\Big\rangle\Big\langle \frac{1}{2}\Big|+\Big|-\frac{1}{2}\Big\rangle\Big\langle -\frac{1}{2}\Big|\Big]\equiv |0,0) \nonumber\\
		&\textrm{Triplet}\qquad\rightarrow\qquad\begin{cases}
			|\xi_{-1}):=\Big|-\frac{1}{2}\Big\rangle\Big\langle \frac{1}{2}\Big|\qquad\qquad\qquad\quad\,\,\equiv |1,-1)\\
			|\xi_0):=\frac{1}{\sqrt{2}}\Big[\Big|\frac{1}{2}\Big\rangle\Big\langle \frac{1}{2}\Big|-\Big|-\frac{1}{2}\Big\rangle\Big\langle -\frac{1}{2}\Big|\Big]\,\,\equiv |1,0)\\
			|\xi_{1}):=-\Big|\frac{1}{2}\Big\rangle\Big\langle -\frac{1}{2}\Big|\qquad\qquad\qquad\qquad\,\equiv |1,1)
		\end{cases}\label{R6}
	\end{align}
	where $| . , . )$ gives the $(n,n_3)$ values. Clearly $r_n^2$ eigenvalue in the triplet subspace can be taken to be the square of the radius for $n=1$ fuzzy $S_*^2$ in units of $\lambda^2: r_1^2=1(1+1)=2$.\\
	Likewise the Hilbert space $\mathcal{H}_F=\mathbb{C}^2\otimes\mathcal{H}_q$ , which is spanned by the Dirac spinors (alternatively the chiral spinors)
	can be shown to split, through Clebsch-Gordon rule, into a quadruplate and a pair of doublet states as- $2\times (2\times 2)=4\oplus 2\oplus 2 $ and will therefore span the representation space of spin- 3/2 and 1/2 fuzzy sphere with $r_{\frac{3}{2}}^2= \frac{3}{2}(\frac{3}{2}+1)\lambda^2$ and $r_{\frac{1}{2}}^2=\frac{1}{2}(\frac{1}{2}+1)\lambda^2$ respectively. That is why, the massive Dirac spinors, which constitute the
	pair of doublets, get assigned with $j_3=\pm \frac{1}{2}$ values, whereas  the 4 massless spinors belonging to the quadruplate, have $j_3$
	values $\{-\frac{3}{2},-\frac{1}{2},\frac{1}{2},\frac{3}{2}\}$ (see  (\ref{B1}) and (\ref{N1}) for reference).
	\section{Representation of algebra generators in $\mathcal{H}_c,\mathcal{H}_q$, and $\mathcal{H}_F$}
	Before taking up the computation of real structure it will be advantageous to construct the representation of algebra generators in the Hilbert spaces $\mathcal{H}_c, \mathcal{H}_q,$ and $\mathcal{H}_F$. \\
	To obtain the representation of algebra in the Hilbert space $\mathcal{H}_c$ we simply left multiply the identity $I_{\mathcal{H}_c}$ (\ref{A2}) by $\hat{x}_1$ and make use of (\ref{A1}) to get,
	\begin{align}
		\hat{x}_1= \hat{x}_1 I_{\mathcal{H}_c}= \hat{x}_1\left(\Big|\frac{1}{2}\Big\rangle\Big\langle \frac{1}{2}\Big|+ \Big|-\frac{1}{2}\Big\rangle\Big\langle -\frac{1}{2}\Big|\right)&= \frac{1}{2}(\hat{x}_++\hat{x}_-)\left(\Big|\frac{1}{2}\Big\rangle\Big\langle \frac{1}{2}\Big|+ \Big|-\frac{1}{2}\Big\rangle\Big\langle -\frac{1}{2}\Big|\right)\nonumber\\
		&=\frac{\lambda}{2}\left(\Big|\frac{1}{2}\Big\rangle\Big\langle -\frac{1}{2}\Big|+\Big|-\frac{1}{2}\Big\rangle\Big\langle \frac{1}{2}\Big|\right)=\frac{\lambda}{2}\sigma_1\label{B2}
	\end{align}
	where the rows and columns of the Pauli matrix $\sigma_1$ are labeled by ($\frac{1}{2}$) and $(-\frac{1}{2})$  respectively.We can proceed  similarly to identify $\hat{x}_2$ and $\hat{x}_3$ as,
	\begin{equation}
		\hat{x}_2= \hat{x}_2 I_{\mathcal{H}_c}= \frac{-i\lambda}{2}\left(\Big|\frac{1}{2}\Big\rangle\Big\langle -\frac{1}{2}\Big|- \Big|-\frac{1}{2}\Big\rangle\Big\langle \frac{1}{2}\Big|\right)=  \frac{\lambda}{2}\sigma_2;\quad
		\hat{x}_3= \hat{x}_3 I_{\mathcal{H}_c}= \frac{\lambda}{2}\left(\Big|\frac{1}{2}\Big\rangle\Big\langle \frac{1}{2}\Big|-\Big|-\frac{1}{2}\Big\rangle\Big\langle -\frac{1}{2}\Big|\right)=\frac{\lambda}{2}\sigma_3 \label{B8}
	\end{equation}
	so that we can combine all these three components to write compactly as.
	\begin{equation}
		\hat{\vec{x}}=\frac{\lambda}{2}\vec{\sigma}\label{N8}
	\end{equation}
	These are the representations of left actions of coordinate operators $\hat{x}_i$ on $\mathcal{H}_c$. The corresponding expressions for the right actions of $\hat{x}_i^R$'s on $\tilde{\mathcal{H}}_c$can be obtained in a similar  manner (as the action of $x_i^R$ is only defined on $\tilde{\mathcal{H}}_c$ ),
	\begin{equation}
		\hat{\vec{x}}^R= \hat{\vec{x}}^RI_{\mathcal{H}_c}= \left(\Big|\frac{1}{2}\Big\rangle\Big\langle \frac{1}{2}\Big|+ \Big|-\frac{1}{2}\Big\rangle\Big\langle -\frac{1}{2}\Big|\right)\hat{\vec{x}}= \frac{\lambda}{2}\vec{\sigma} \label{B5}
	\end{equation}
	Note that both the left and right actions on $\mathcal{H}_c$ and $\tilde{\mathcal{H}}_c$ have apparently the same forms in terms of Pauli matrices (\ref{N8},\ref{B5}) but their domain of actions remain different; it is $\mathcal{H}_c$ for $\hat{x}_i$ and $\tilde{\mathcal{H}}_c$ for $\hat{x}_i^R$. It therefore becomes quite obvious that $\vec{\hat{x}}^R$ s satisfies the condition of opposite algebra i.e. 
	\begin{equation}
		[\hat{x}_i,\hat{x}_j^R]=0;\,\,\,\,\, (x_i^Rx_j^R)=x_j^Rx_i^R \label{N3}
	\end{equation} 
	and therefore  the lie- algebra satisfied by $\hat{x}_i^R$ picks up a minus sign in its structure constant:
	\begin{equation}
		[\hat{x}_i^R,\hat{x}_j^R]= -i\lambda\epsilon_{ijk}\hat{x}_k^R \label{N5}
	\end{equation}
	Now we will represent the algebra generators in the basis of $\mathcal{H}_q$. To this end, let us first denote the four basis elements of $\mathcal{H}_q$ in (\ref{eq2}) as
	\begin{align}
		|\eta_1 ) &= \Big|\frac{1}{2}\Big\rangle\Big\langle \frac{1}{2}\Big|=\Big|\frac{1}{2}, \frac{1}{2}\Big) ;\,\,\,\,|\eta_2 )=\Big|\frac{1}{2}\Big\rangle\Big\langle -\frac{1}{2}\Big|= \Big|\frac{1}{2},-\frac{1}{2}\Big);\nonumber\\ |\eta_3)& =\Big|-\frac{1}{2}\Big\rangle\Big\langle \frac{1}{2}\Big|= \Big|-\frac{1}{2},\frac{1}{2}\Big) ;\,\,\,\,\,|\eta_4) =\Big|-\frac{1}{2}\Big\rangle\Big\langle -\frac{1}{2}\Big|= \Big|-\frac{1}{2},-\frac{1}{2}\Big) \label{eq46}
	\end{align}  
	so that the completeness relations (\ref{A3}) for $\mathcal{H}_q$  can be written  as 
	\begin{equation}
		\sum_{\mu=1}^4 |\eta_{\mu})(\eta_{\mu} | = \textbf{I}_{\mathcal{H}_q};\,\,\,\,\,\,\mu=1,2,3,4 \label{eq47}
	\end{equation}
	At this point it will be convenient to express the basis $|\eta_{\mu})$ in terms of the algebra generators so that we can use them in our future calculations.
	\begin{equation}
		|\eta_1) =\begin{pmatrix}
			1&0\\ 0&0
		\end{pmatrix}= \frac{\textbf{1}}{2}+\frac{\hat{x}_3}{\lambda};\,\,|\eta_2) = \begin{pmatrix}
			0&1\\0&0
		\end{pmatrix}=\frac{\hat{x}_+}{\lambda};\,\,\,|\eta_3)=\begin{pmatrix}
			0&0\\1&0
		\end{pmatrix}=\frac{\hat{x}_-}{\lambda};\,\,\,|\eta_4)=\begin{pmatrix}
			0&0\\0&1
		\end{pmatrix}=\frac{\textbf{1}}{2}-\frac{\hat{x}_3}{\lambda}\label{N51}
	\end{equation}
	We can now determine representation of $\hat{x}_i$ on $\mathcal{H}_q$  by just left multiplying $\textbf{I}_{\mathcal{H}_q}$ by $\hat{x}_i$ as we did in (\ref{B2}) to get after a straightforward computation, 
	\begin{equation}
		\hat{x}_j=\hat{x}_j I_{\mathcal{H}_q}= \sum_{\mu=1}^4 |\hat{x}_j\eta_{\mu}) (\eta_{\mu}| =\frac{\lambda}{2} \sigma_j \otimes I_2 \label{N10}
	\end{equation}
	And for the right action $\hat{x}_j^R$ we can like-wise obtain
	\begin{equation}
		\hat{x}_j^R= \hat{x}_j^R I_{\mathcal{H}_q}=\sum_{\mu = 1}^4 |\eta_{\mu}\hat{x}_j)(\eta_{\mu}| =\frac{\lambda}{2} I_2\otimes \sigma_j \label{N11}
	\end{equation}
	so that we can write both (\ref{N10},\ref{N11}) in a more compact index-free notation as 
	\begin{equation}
		\hat{\vec{x}}=\frac{\lambda}{2} \vec{\sigma}\otimes I_2;\,\,\,\,\,\,\hat{\vec{x}}^R=\frac{\lambda}{2}I_2\otimes \vec{\sigma}\label{N12}
	\end{equation}
	Note that, in (\ref{N10}, \ref{N11}) the operator $\hat{x}_i$ acts on $|\eta_{\mu})$ from left and right respectively and does not touch the $(\eta_{\mu}|$ sector. In addition the operators (\ref{N10}, \ref{N11}) act on $\mathcal{H}_q$, where the first entry of the tensor product acts on $\mathcal{H}_q$ from left and the second entry acts on $\mathcal{H}_q$ from right. One can now easily verify that both $\hat{\vec{x}}$ and $\hat{\vec{x}}^R$ satisfy their respective su(2) and opposite su(2)$^R$  algebras \footnote{$x_i^Rx_j^R |m,n) = \frac{\lambda^2}{4} (I_2\otimes\sigma_i)(I_2\otimes\sigma_j)(|m\rangle\langle n|)=\frac{\lambda^2}{4}(I_2\otimes\sigma_i)(|m\rangle\langle n| \sigma_j) =\frac{\lambda^2}{4}|m\rangle\langle n|\sigma_j\sigma_i. $}, which are isomorphic to (\ref{eq1}) and (\ref{N5}) respectively.\\  
	Finally, we need to find the representation of $\hat{x}_i$ on $\mathcal{H}_F=\mathbb{C}^2 \otimes \mathcal{H}_q$.   $\hat{x}_i$ acts on $\mathcal{H}_F$ through the diagonal representation $\pi$ : \begin{equation}
		\pi(a)=\begin{pmatrix}
			a&0\\0&a
		\end{pmatrix}
		;\,\,\,\,\,\,\,\,\pi(a^R)=\begin{pmatrix}a^R&0\\0&a^R\end{pmatrix};\,\,\,\,\, a\in \mathcal{A}_F=\mathcal{H}_q\label{B6}
	\end{equation} 
	The completeness condition for the space $\mathcal{H}_F$ can be written more explicitly as,
	\begin{equation}
		\sum_{\mu=1}^4 \left[ \begin{pmatrix}
			\eta_{\mu}\\0
		\end{pmatrix} (\eta_{\mu}^* \,\,\,\,\,\,0) + \begin{pmatrix}
			0\\\eta_{\mu} \end{pmatrix}(0 \,\,\,\, \eta_{\mu}^*)\right] =\textbf{I}_{\mathcal{H}_F} \label{eq48}
	\end{equation}
	So we can write, in particular, for $\hat{x}_1$ as, 
	\begin{equation}
		\pi(\hat{x}_1) = \pi (\hat{x}_1) I_{\mathcal{H}_F}  =\sum _{\mu=1}^4 \left[\begin{pmatrix}
			\hat{x}_1 \eta_{\mu}\\ 0
		\end{pmatrix}(\eta_{\mu}^* \,\,\,\,\,0)+ \begin{pmatrix}
			0\\ \hat{x}_1 \eta_{\mu}
		\end{pmatrix}(0\,\,\,\,\, \eta_{\mu}^*) \right] \label{eq49}
	\end{equation}
	A straightforward computation using (\ref{N2}, \ref{eq46}) and 
	\begin{equation}
		\hat{x}_1 \eta_1=\frac{\lambda}{2}\eta_3;\,\,\,\,\,\hat{x}_1 \eta_2=\frac{\lambda}{2}\eta_4;\,\,\,\,\hat{x}_1 \eta_3=\frac{\lambda}{2}\eta_1;\,\,\,\,\hat{x}_1 \eta_4=\frac{\lambda}{2}\eta_2 \label{N49}
	\end{equation} 
	yields 
	\begin{equation}
		\pi(\hat{x}_1)= \frac{\lambda}{2}\left[|\phi_3)) \,(( \phi_1| +|\phi_4)) \,(( \phi_2| +|\phi_1)) \,(( \phi_3| +|\phi_2)) \,(( \phi_4|+|\phi_7)) \,(( \phi_5|+|\phi_8)) \,(( \phi_6|+|\phi_5)) \,(( \phi_7|+ |\phi_6)) \,(( \phi_8|\right] \label{eq51}
	\end{equation}
	This can be recast in the form of a matrix as 
	\begin{equation}
		\pi(\hat{x}_1)=\frac{\lambda}{2}\begin{pmatrix}
			0&0&1&0& & & & \\
			0&0&0&1& & & &\\
			1&0&0&0& & & &\\
			0&1&0&0& & & &\\
			& & & &0&0&1&0\\
			& & & &0&0&0&1\\
			& & & &1&0&0&0\\
			& & & &0&1&0&0
		\end{pmatrix}=\frac{\lambda}{2}
		\begin{pmatrix}
			\sigma_1\otimes \textbf{I}_2 &0\\
			0& \sigma_1 \otimes \textbf{I}_2
		\end{pmatrix}=\frac{\lambda}{2} \textbf{I}_2\otimes (\sigma_1\otimes \textbf{I}_2) \label{eq52}
	\end{equation}
	where the rows and columns are labelled by $\phi_1,\phi_2,....,\phi_8$ in this order.\\
	Proceeding like-wise we get similar structure for $\pi(\hat{x}_2)$ and $\pi(\hat{x}_3)$ , and enables us to write all these more compactly as
	\begin{equation}
		\pi(\vec{x})=\frac{\lambda}{2} \textbf{I}_2 \otimes (\vec{\sigma}\otimes \textbf{I}_2) \label{eq53}
	\end{equation}
	Coming to the right action $\pi(\hat{\vec{x}}^R)$, we again consider the first component $\pi(\hat{x}_1^R)$, which can be written like (\ref{eq49}) as, 
	\begin{equation*}
		\pi(\hat{x}_1^R) = \pi(\hat{x}_1^R)\textbf{I}_{\mathcal{H}_F} =\sum_{i=1}^4 \begin{pmatrix}
			\hat{x}_1^R&0\\0&\hat{x}_1^R
		\end{pmatrix}\left[\begin{pmatrix}
			\eta_i\\0
		\end{pmatrix}(\eta_i^*\,\,\,\,\, 0)+ \begin{pmatrix}
			0\\\eta_i
		\end{pmatrix} (0 \,\,\,\,\eta_i) \right]
	\end{equation*}
	\begin{equation}
		=\sum_{i=1}^4 \left[\begin{pmatrix}
			\eta_i\hat{x}_1\\0
		\end{pmatrix}(\eta_i^*\,\,\,\,\, 0)+ \begin{pmatrix}
			0\\\eta_i\hat{x}_1
		\end{pmatrix} (0 \,\,\,\,\eta_i) \right] \label{eq54}
	\end{equation}
	Now making use of the identities 
	\begin{equation}
		\eta_1\hat{x}_1= \frac{\lambda}{2} \eta_2;\,\,\,\,\,\,\,\,\,\eta_2\hat{x}_1= \frac{\lambda}{2} \eta_1;\,\,\,\,\,\,  \eta_3\hat{x}_1= \frac{\lambda}{2} \eta_4;\,\,\,\,\,\,
		\eta_4\hat{x}_1= \frac{\lambda}{2} \eta_3, \label{eq55}
	\end{equation}
	we readily obtain 
	\begin{equation}
		\pi(\hat{x}_1^R)= \frac{\lambda}{2}\left[|\phi_2)) \,(( \phi_1| +|\phi_1)) \,(( \phi_2| +|\phi_4)) \,(( \phi_3| +|\phi_3)) \,(( \phi_4|+|\phi_5)) \,(( \phi_6|+|\phi_6)) \,(( \phi_5|+|\phi_7)) \,(( \phi_8|+ |\phi_8)) \,(( \phi_7|\right] \label{eq56}
	\end{equation}
	Again its matrix form can virtually read off as,
	\begin{equation}
		\pi(\hat{x}_1^R)=\frac{\lambda}{2}\begin{pmatrix}
			0&1&0&0& & & & \\
			1&0&0&0& & & &\\
			0&0&0&1& & & &\\
			0&0&1&0& & & &\\
			& & & &0&1&0&0\\
			& & & &1&0&0&0\\
			& & & &0&0&0&1\\
			& & & &0&0&1&0
		\end{pmatrix}=\frac{\lambda}{2}
		\begin{pmatrix}
			\textbf{I}_2\otimes\sigma_1 &0\\
			0& \textbf{I}_2\otimes\sigma_1
		\end{pmatrix}=\frac{\lambda}{2} \textbf{I}_2\otimes ( \textbf{I}_2\otimes \sigma_1) \label{eq57}
	\end{equation}
	And proceeding similarly we arrive at the corresponding representation of the right action in a more compact form as 
	\begin{equation}
		\pi(\hat{\vec{x}}^R)=\frac{\lambda}{2}\textbf{1}_2 \otimes (\textbf{1}_2\otimes \vec{\sigma})\label{N9}
	\end{equation}
	Now from the very structures of $\pi(\hat{\vec{x}})$ (\ref{eq53})  and $\pi(\hat{\vec{x}}^R)$ (\ref{N9}), it is clear that 
	\begin{equation}
		[\pi(\hat{x}_i), \pi(\hat{x}_j^R)]= 0\,\,\,\,\,\,\,\,\,\,\forall \,\,\,i,j\,\in \{1,2,3\}\label{eq59}
	\end{equation}
	and
	\begin{equation}
		[\pi(\hat{x}_i^R),\pi(\hat{x}_j^R)]=-i\lambda \epsilon_{ijk}\pi(\hat{x}_k^R)\label{eq60}
	\end{equation}
	\section{Determination of real structure $\mathcal{J}_F$}
	Now since we have the grading operator $\gamma_F$ already existing (\ref{eq8}), the internal spectral triple must have even KO dimension. It therefore follows immediately that the real structure $\mathcal{J}_F$ , if it also exists , must commute with the Dirac operator $D_F$ (\ref{eq8})  \cite{suij,barr}:
	\begin{equation}
		[D_F,\mathcal{J}_F]=0 \label{eq21}
	\end{equation}   
	Therefore , this operator should connect states having the same mass eigen values . For $m=\pm 1$, there are virtually no ambiguities and one can safely take
	\begin{equation}
		|\pm 1, \frac{1}{2})) \,\,\,\,\longleftrightarrow \,\,\,\, \pm |\pm 1, -\frac{1}{2})) \label{eq22}
	\end{equation}  
	Further in the paper \cite{barr} the author has indicated that the KO dimension of a triple should be 0 or 4 if $I$ defined below is non-zero:
	\begin{equation}
		I= dim(H_+)-dim(H_-)\ne 0
	\end{equation}
	where $H_+$ is a subspace of the total Hilbert space $\mathcal{H}_F$ which is also the positive eigen-space of the chirality operator $\gamma_F$ and $H_-$ is the negative chirality subspace. As can be seen in our case, using  (\ref{eq19}) and (\ref{eq20})  that $I=$2-6=-4$\ne 0$. So we can restrict our search for  KO dim = 0 and 4.\\
	Writing more compactly (i.e suppressing the double ket notation used in (\ref{B1})), we can have either,
	\begin{align}
		&\mathcal{J}_F \psi_1 =\psi_2;\,\,\,\,\mathcal{J}_F \psi_3 =\psi_4;\,\,\,\,\,\,\,\mathcal{J}_F\psi_2=\psi_1;\,\,\,\,\,\,\,\mathcal{J}_F\psi_4=\psi_3 \label{eq23}\\
		\textrm{or}\qquad\qquad
		&\mathcal{J}_F \psi_1 =\psi_2;\,\,\,\,\mathcal{J}_F \psi_3 = \psi_4\,\,\,\,\,\,\,\mathcal{J}_F\psi_2=-\psi_1;\,\,\,\,\,\,\,\mathcal{J}_F\psi_4=-\psi_3\label{eq24}
	\end{align}
	Considering these possibilities , we see immediately from the structures of $\psi_1,....,\psi_8$ (\ref{B1}), that the simple swapping  $\Big\langle\frac{1}{2}\Big| \longleftrightarrow \Big\langle  -\frac{1}{2}\Big|$ and $\Big|\frac{1}{2}\Big \rangle \longleftrightarrow \Big|-\frac{1}{2}\Big\rangle$ followed by an interchanging of upper and lower components of all the eight spinors $\psi_1,...,\psi_8$ in (\ref{B1}) accomplishes the transformation in (\ref{eq23},\ref{eq24}) and extends even to the massless sector as,
	\begin{align}
		&\mathcal{J}_F \psi_5=\psi_6;\,\,\,\,\,\,\mathcal{J}_F \psi_7=\psi_8;\,\,\,\,\,\mathcal{J}_F\psi_6=\psi_5;\,\,\,\,\,\,\mathcal{J}_F\psi_8=\psi_7 \label{A5}\\
		\textrm{or}\qquad\qquad
		&\mathcal{J}_F \psi_5=\psi_6;\,\,\,\,\,\,\mathcal{J}_F \psi_7=\psi_8;\,\,\,\,\,\mathcal{J}_F\psi_6=-\psi_5;\,\,\,\,\,\,\mathcal{J}_F\psi_8=-\psi_7 \label{A9}
	\end{align}
	The above swapping operation in $\mathcal{H}_c$ implies the following swapping operation at the level of $\mathcal{H}_q$:
	\begin{equation}
		\Big|\frac{1}{2}\Big\rangle\Big\langle\frac{1}{2}\Big|\longleftrightarrow \Big|-\frac{1}{2}\Big\rangle\Big\langle -\frac{1}{2}\Big|;\,\,\,\, \Big|\frac{1}{2}\Big\rangle\Big\langle -\frac{1}{2}\Big|\longleftrightarrow\Big|-\frac{1}{2}\Big\rangle\Big\langle\frac{1}{2}\Big| \label{eq25}
	\end{equation}
	Finally at the level of the total Hilbert space $\mathcal{H}_F=\mathcal{H}_q\otimes\mathbb{C}^2$, this implies the swapping within the following four pairs of canonical basis vectors (\ref{N2}) :
	\begin{equation}
		\phi_1\longleftrightarrow \phi_8;\,\,\,\,\phi_2\longleftrightarrow \phi_7;\,\,\,\, \phi_3\longleftrightarrow\phi_6;\,\,\,\,\phi_4\longleftrightarrow\phi_5\label{eq26}
	\end{equation} 
	The signs can be different depending on the KO dimension 0 or 4.
	Note that here we had to augment the swapping in (\ref{eq25}) with the exchange of upper and lower components. The importance of real structure, if it exists, should enable us to relate the elements of the opposite algebra $\mathcal{A}_F^o$ to that of $\mathcal{A}_F$ as,
	\begin{equation}
		\pi(a^o)=\mathcal{J}_F \pi(a^*)\mathcal{J}_F^*;\,\,\,\,\,\,\,\,\,\,\,\,\,a\in \mathcal{A}_F=\mathcal{H}_q, \textrm{and} \,\,\,\,\,\,a^o\,\,\in \mathcal{A}_F^o =\mathcal{H}_q^o \label{N4}\end{equation}
	fulfilling \begin{equation}
		[\pi(a),\pi(b^o)]=0,\,\,\,\,\,\,\,\,\, \forall a,b \in \mathcal{A}_F\,\,\,\,\,\,\,(\textrm{zero order condition})\label{eq29}
	\end{equation}
	and, preferably, also 
	\begin{equation}
		[[D_F,\pi(a)],\pi(b^o)]=0 \,\,\,\,\,\,\,\,\,\,\,\,\,\,\,\,(\textrm{first order condition})\label{eq30}
	\end{equation}
	For this, one may expect to identify $\hat{a}^o = \hat{a}^R $ eventually, so that the opposite algebra $\mathcal{A}_F^o$ can be thought of being generated by $\hat{x}_i^R$.\\
	As it turns out there are no consistent solution for real structure $\mathcal{J}_F$ satisfying the properties appropriate for  KO dimension=0 and (\ref{N4}).	So we consider the transformations 	(\ref{eq24}) and (\ref{A9}) and try to construct a real structure operator for KO dimension= 4.\\
	The Hilbert space $\mathcal{H}_q$ can be splitted into two sub-spaces as $\mathcal{H}_q=\mathcal{H}_q^{L_3=0} \oplus \mathcal{H}_q^{L_3\ne 0}$, the former with the eigen states $L_3=0$ (i.e linear span of $\eta_1$,$\eta_4$) and later with eigen states having $\L_3=\pm 1$ (i.e. linear span of $\eta_2$ and $\eta_3$). 
	(\ref{eq46}) indicates that the action of $\mathcal{J}_F$ should be such that it swaps between the eigen spinors in the respective sectors, also interchanging the components of the $\mathbb{C}^2$that is the lower and upper components of the spinors. To show the transformation explicitly sector-wise we proceed as follows :\\
	(1)\begin{equation}
		L_3=0\,\, \textrm{Sector}: \begin{pmatrix}\eta_1\\0\end{pmatrix}   \rightarrow \begin{pmatrix}
			0\\ -\eta_4\end{pmatrix};\,\,\,\begin{pmatrix}\eta_4\\0\end{pmatrix}   \rightarrow \begin{pmatrix}
			0\\ -\eta_1\end{pmatrix};\,\,\,\begin{pmatrix}0\\\eta_1\end{pmatrix}   \rightarrow \begin{pmatrix}
			\eta_4\\ 0\end{pmatrix};\,\,\,\begin{pmatrix}0\\\eta_4\end{pmatrix}   \rightarrow \begin{pmatrix}
			\eta_1\\0 \end{pmatrix} 
	\end{equation}  
	The interchange of $\eta_1$ and $\eta_4$ implies the following conversion in $\mathcal{H}_q$ (we have used ( \ref{N51})):
	\begin{equation}
		\hat{x}_3 \to -\hat{x}_3
	\end{equation}
	(2) \begin{equation}
		L_3=\pm 1\,\, \textrm{Sector}: \begin{pmatrix}\eta_3\\0\end{pmatrix} \rightarrow \begin{pmatrix}0\\\eta_2\end{pmatrix};\,\,\,\begin{pmatrix}0\\\eta_2\end{pmatrix}\rightarrow \begin{pmatrix}-\eta_3\\0\end{pmatrix}   
	\end{equation}
	Here  $\eta_2 \leftrightarrow \eta_3$ implies the  following interchanges in $\mathcal{H}_q$:
	\begin{equation}
		\hat{x}_+\,\,\, \leftrightarrow \,\,\,\hat{x}_-\label{N46}
	\end{equation}
	With the above interchanges in mind it is tempting to try constructing the real structure by including a total space inversion followed by a hermitian conjugation.\\
	Let us first try to implement the space inversion with, 
	\begin{equation}
		\hat{\vec{x}}\to \hat{x}^{\prime} \to P\hat{\vec{x}} = - \hat{\vec{x}}\label{N14},
	\end{equation}
	where the parity operator $P$ can be unitarily implemented  in $\mathcal{H}_q$ by the following transformation in the $\eta_{\mu}$ basis (\ref{eq46}) as,
	\begin{equation}
		\begin{pmatrix}
			\eta_1\\ \eta_2\\ \eta_3\\ \eta_4
		\end{pmatrix} \to  \begin{pmatrix}
			\eta_1^{\prime}\\ \eta_2^{\prime}\\ \eta_3^{\prime}\\ \eta_4^{\prime}
		\end{pmatrix}= \begin{pmatrix}
			\frac{4}{\lambda^2}\hat{x}_1\hat{x}_1^R&0&0&0\\
			0&e^{i\pi\hat{L}_3}&0&0\\
			0&0& e^{i\pi\hat{L}_3}&0\\
			0&0&0&\frac{4}{\lambda^2}\hat{x}_1\hat{x}_1^R
		\end{pmatrix} 	\begin{pmatrix}
			\eta_1\\ \eta_2\\ \eta_3\\ \eta_4
		\end{pmatrix}=\begin{pmatrix}
			0&0&0&1\\0&-1&0&0\\0&0&-1&0\\1&0&0&0
		\end{pmatrix} 	\begin{pmatrix}
			\eta_1\\ \eta_2\\ \eta_3\\ \eta_4
		\end{pmatrix}\label{N25}
	\end{equation}
	so that the following relations
	\begin{equation}
		\eta_1^{\prime}=P\eta_1 =\eta_4;\,\,\eta_2^{\prime} =P\eta_2=-\eta_2;\,\,\eta_3^{\prime}=P\eta_3 =-\eta_3;\,\,\eta_4^{\prime}=P\eta_4= \eta_1 \label{N26}	
	\end{equation}
	hold, where the transformation  (\ref{N14}) becomes quite obvious if we use (\ref{N26}).\\
	We should mention at this stage , however, that under the transformation (\ref{N14}) the original commutator bracket (1) is not invariant as it stands, as it satisfies only an SO(3) symmetry. In other words, the commutation relation will is not preserved if its structure constant  $\epsilon_{ijk}$ is treated as a pseudo tensor and held fixed under parity transformation. We can however make it to respect the enlarged O(3) symmetry as well , if we elevate $\epsilon_{ijk}$ to $E_{ijk}$  which is now a tensor satisfying the  following property,
	\begin{align*}
		E_{ijk} &= \epsilon_{ijk},\,\,\,\,\textrm{for right handed system} \\
		&=-\epsilon_{ijk}\,\,\,\,\textrm{for left handed system}
	\end{align*}
	where $\epsilon_{123}=+1$ as is usually defined in a typical right handed system. In this context we would like to mention that for our $n=\frac{1}{2}$ case, the unital algebra $\mathcal{A}_F$ (4), which can also be expressed as 
	\begin{equation}
		\mathcal{A}_F=Span_{\mathbb{C}}\{\eta_1,\eta_2,\eta_3,\eta_4\} = Span_{\mathbb{C}}\{I_2, \hat{x}_1,\hat{x}_2,\hat{x}_3\}  \label{N29}
	\end{equation} 
	follows the multiplication rule:
	\begin{equation}
		\hat{x}_i\hat{x}_j =\frac{\lambda^2}{4}\delta_{ij}I_2 +\frac{i\lambda}{2}\epsilon_{ijk} \hat{x}_k \,\,\,\,\in \,\,\,\mathcal{A}_F\label{N30}
	\end{equation}
	appropriate for a right handed system. 
	Particularly for $i \ne j, $  in a right handed system,
	$$\hat{x}_1\hat{x}_2 =\frac{i\lambda}{2} \hat{x}_3 ,\,\,\,\,\textrm{and cyclic permutation}$$
	while for the left handed system 
	\begin{equation}
		\hat{x}_1\hat{x}_2 =-\frac{i\lambda}{2} \hat{x}_3,\,\,\,\,\textrm{and cyclic permutation}
	\end{equation}
	Note that left handed system can be obtained from the right handed system by the action of any generic O(3) element satisfying det=-1. The above parity operator introduced in (\ref{N14}) is one such example. 
	So generally  for a left-handed system the multiplication rule can be defined as,
	\begin{equation}
		\hat{x}_i\hat{x}_j =\frac{\lambda^2}{4}\delta_{ij}I_2 -\frac{i\lambda}{2}\epsilon_{ijk} \hat{x}_k \,\,\,\,\in \,\,\,\mathcal{A}_F\label{N52}
	\end{equation}
	So we propose a new multiplication rule which combines those of both right handed (\ref{N30}) and left-handed (\ref{N52}) systems and is given by,
	\begin{equation}
		\hat{x}_i\hat{x}_j =\frac{\lambda^2}{4}\delta_{ij}I_2 +\frac{i\lambda}{2}E_{ijk} \hat{x}_k \,\,\,\,\in \,\,\,\mathcal{A}_F\label{N53}
	\end{equation}
	This ensures,
	\begin{equation}
		P(\hat{x}_i\hat{x}_j) = \hat{x}_i\hat{x}_j \label{N57}
	\end{equation} 
	So the multiplication rule is now changed, keeping the  underlying vector space of the algebra same. Now the coordinate algebra (\ref{eq1}) remains invariant under the action of $P$, and the total symmetry group is now enhanced to O(3) from SO(3) as mentioned earlier.
	\footnote{We would like to mention that the relation (\ref{N30}) is more stringent than (1) in the sense that (1) follows from (\ref{N30}) by simple anti-symmetrization, but the converse is not true; it will not hold for fuzzy $S^2_*$ associated with any $n\ge 1$ representation. For example, n=1 case, one can take $(\rho_1(\hat{x}_i))_{jk}=-i\lambda\epsilon_{ijk}$ fulfilling (1), but will not satisfy (\ref{N30}).}Infact the O(3) symmetric Lie algebra satisfied by $\hat{x}_i$'s now is simply obtained by anti-symmetrizing (\ref{N53}) to get 
	\begin{equation}
		[\hat{x}_i,\hat{x}_j]=i\lambda E_{ijk} \hat{x}_k\label{N54}
	\end{equation}
	The corresponding Lie-algebra satisfied by the right action $\hat{x}_i^R$ is therefore   given by, 
	\begin{equation}
		[\hat{x}_i^R,\hat{x}_j^R]=-i\lambda E_{ijk} \hat{x}_k^R\label{N55}	
	\end{equation}
	which too enjoys an enlarged O(3) symmetry now. The $O(3)$ covariant fuzzy sphere has been investigated earlier in \cite{fio}, where it was shown that, a O(3) covariant fuzzy sphere is built by imposing a energy cutoff on a quantum particle confined in a potential of the form $V(r)$.\\
	It therefore becomes quite obvious that this parity symmetry too should be implemented through an automorphism symmetry of the algebra $\mathcal{A}_F$ (4) . And this can be done by requiring the elements of $\mathcal{A}_F$ to transform as scalar under parity (\ref{N14}):
	\begin{equation}
		a(I_2,\hat{\vec{x}}) \to a^{\prime} (I_2,\hat{\vec{x}}^{\prime}) = a(I_2,\hat{\vec{x}});\,\,\,\,\,\hat{\vec{x}}^{\prime}:=-\hat{\vec{x}} \label{N31}
	\end{equation}
	Next we can implement the transformation (\ref{N46}) in $\mathcal{H}_q$ by a hermitian conjugation $H$  (the involution operator) or $*$ operation introduced below (\ref{N44}).
	Further, we need an operator which can act on the $\mathbb{C}^2$ sector of the total Hilbert space and interchanges the upper and lower components.  \\
	Having all the basic ingredients in our disposal, we can now introduce the real structure as,
	\begin{equation}
		\mathcal{J}_F= (i\sigma_2)\otimes (H \circ P) = \begin{pmatrix}
			0&1\\-1&0
		\end{pmatrix} \otimes (H\circ P) \label{N32}
	\end{equation} 
	where P (\ref{N14},\ref{N25}) acts linearly. It can be further checked that $\mathcal{J}_F$ as defined in (\ref{N32}) is an anti-unitary and anti-linear operator. We now, as an example, demonstrate $\mathcal{J}_F\psi_1 = -\psi_2$, where $\psi_1$ and $\psi_2$ are eigen spinors (\ref{B1}) of the Dirac operator $\mathcal{D}_F$ . For that, it will be convenient to first re-express $\psi_1$, $\psi_2$ in terms of \{$\eta_{\mu}$\} to get,
	\begin{equation}
		\psi_1=N\begin{pmatrix}
			\eta_1 +(\sqrt{3}-1)\eta_4\\(2-\sqrt{3})\eta_2
		\end{pmatrix} \textrm{and} \,\,\,\psi_2=N\begin{pmatrix}
			(2-\sqrt{3})\eta_3\\ \eta_4+(\sqrt{3}-1)\eta_1
		\end{pmatrix};\,\,\,\, N=\frac{1}{3-\sqrt{3}}\label{N33}
	\end{equation}
	Then
	\begin{equation}
		\mathcal{J}_F\psi_1=N\begin{pmatrix}
			0&1\\-1&0
		\end{pmatrix} \begin{pmatrix}
			(H\circ P)(\eta_1 +(\sqrt{3}-1)\eta_4)\\(H\circ P)(2-\sqrt{3})\eta_2
		\end{pmatrix}= N\begin{pmatrix}
			0&1\\-1&0
		\end{pmatrix}\begin{pmatrix}
			\eta_4+(\sqrt{3}-1)\eta_1 \\ -(2-\sqrt{3}) \eta_3\end{pmatrix}
		=-N\begin{pmatrix}
			(2-\sqrt{3}) \eta_3\\\eta_4+(\sqrt{3}-1)\eta_1 
		\end{pmatrix}=-\psi_2 \label{N34} 
	\end{equation}
	where we have made use of (\ref{N8}) and the following set of relations:
	\begin{equation}
		(H\circ P) \eta_1 =\eta_4;\,\,\,(H\circ P) \eta_2 =-\eta_3;\,\,\,(H\circ P) \eta_3 =-\eta_2;\,\,\,(H\circ P) \eta_4 =\eta_1 \label{N35}
	\end{equation}
	Proceeding in a similar manner, we get
	\begin{equation}
		\mathcal{J}_F\psi_2=\psi_1;\,\,\,\mathcal{J}_F\psi_3=-\psi_4;\,\,\,\mathcal{J}_F\psi_4=\psi_3;\,\,\,\mathcal{J}_F\psi_5=\psi_6;\,\,\,\mathcal{J}_F\psi_6=-\psi_5;\,\,\,\mathcal{J}_F\psi_7=-\psi_8;\,\,\,\mathcal{J}_F\psi_8=\psi_7;\,\,\,\label{N36}
	\end{equation} 
	This implies 
	\begin{equation}
		\mathcal{J}_F^2=-1\label{N37}
	\end{equation}
	appropriate for KO dimension-4. We are thus left with the task of verifying (\ref{N4}) with real structure . To that end, consider $(H\circ P)(\hat{x}_i\eta_1)$ for example. Upon simplification, using ( \ref{N30},\ref{N57}), this can be recast in the form 
	\begin{equation}
		(H\circ P)(\hat{x}_i\eta_1)=(H\circ P)(\frac{\hat{x}_i}{2}+\frac{\hat{x}_i\hat{x}_3}{\lambda})=H (-\frac{\hat{x}_i}{2}+\frac{\hat{x}_i\hat{x}_3}{\lambda})=-\frac{\hat{x}_i}{2}+\frac{\hat{x}_3\hat{x}_i}{\lambda}=-\hat{x}_i^R\eta_4 \label{N38}
	\end{equation} 
	Similarly we get for other components of $\eta_{\mu}$ the following relations,
	\begin{equation}
		(H\circ P) (\hat{x}_i\eta_2) = \hat{x}_i^R\eta_3;\,\,\,(H\circ P)(\hat{x}_i\eta_3)= \hat{x}_i^R\eta_2;\,\,\,(H\circ P)(\hat{x}_i\eta_4)=-\hat{x}_i^R\eta_1\label{N39}
	\end{equation}
	Eventually, we get the following relations, using (\ref{N35}), giving the actions of $\mathcal{J}_F$ on the canonical basis (\ref{N2}) of $\mathcal{H}_F$:
	\begin{align}
		&\mathcal{J}_F|\phi_1)) = -|\phi_8));\,\,\, \mathcal{J}_F|\phi_2))=|\phi_7));\,\,\,\mathcal{J}_F|\phi_3))=|\phi_6));\,\,\,\mathcal{J}_F|\phi_4))=-|\phi_5))\nonumber\\
		& \mathcal{J}_F|\phi_5))=|\phi_4));\,\,\,\mathcal{J}_F|\phi_6))=-|\phi_3));\,\,\,\mathcal{J}_F|\phi_7))=-|\phi_2));\,\,\,\mathcal{J}_F|\phi_8))=|\phi_1))\label{N42}
	\end{align}
	Using all these one can write down using (\ref{eq49}),
	\begin{align}
		\mathcal{J}_F\pi(\hat{x}_1)^{\dagger}\mathcal{J}_F^{\dagger}=\mathcal{J}_F\pi(\hat{x}_1)\textbf{1}_{\mathcal{H_F}}\mathcal{J}_F^{\dagger}=	\mathcal{J}_F\sum_{i=1}^8 \pi(\hat{x}_1)|\phi_i)) \,((\phi_i|  \,\mathcal{J}_F^{\dagger}\label{N50}
	\end{align}
	For the purpose of explicit demonstration, let us work with first term from the, as a sample, right hand side of the above equation given by,
	\begin{align}
		\mathcal{J}_F \pi(\hat{x}_1) |\phi_1)) \,((\phi_1|\mathcal{J}_F^{\dagger}&= 	-\mathcal{J}_F\begin{pmatrix}
			\hat{x}_1\eta_1\\0
		\end{pmatrix}	\,((\phi_8|\nonumber\\
		&= -\begin{pmatrix}
			0&1\\-1&0
		\end{pmatrix}\otimes (H\circ P) \begin{pmatrix}
			\hat{x}_1(\frac{1}{2}+\frac{\hat{x}_3}{\lambda})\\0
		\end{pmatrix}\,((\phi_8|\nonumber\\
		&=-\begin{pmatrix}
			0&1\\-1&0
		\end{pmatrix}\otimes H \begin{pmatrix}
			\frac{-\hat{x}_1}{2}+\frac{\hat{x}_1\hat{x}_3}{\lambda}\\0
		\end{pmatrix}\,((\phi_8|\nonumber\\
		&=-\begin{pmatrix}
			0\\ \frac{\hat{x}_1}{2}-\frac{\hat{x}_3\hat{x}_1}{\lambda}
		\end{pmatrix}\,((\phi_8|= -\begin{pmatrix}
			0\\ \eta_4\hat{x}_1
		\end{pmatrix} ((\phi_8|= -|\phi_7)) \,(( \phi_8|
	\end{align}
	Proceeding similarly for every term in the right hand side of (\ref{N50}), we get,
	\begin{align}
		&	\mathcal{J}_F\pi(\hat{x}_1)^{\dagger}\mathcal{J}_F^{\dagger}=  -\frac{\lambda}{2}\left[|\phi_2)) \,(( \phi_1| +|\phi_1)) \,(( \phi_2| +|\phi_4)) \,(( \phi_3| +|\phi_3)) \,(( \phi_4|\right.\nonumber\\
		&\left.+|\phi_5)) \,(( \phi_6|+|\phi_6)) \,(( \phi_5|+|\phi_7)) \,(( \phi_8|+ |\phi_8)) \,(( \phi_7|\right]= -\pi(\hat{x}_1^R)
	\end{align}
	For all the algebra generators we can readily show
	\begin{equation}
		\pi(\hat{x}_i^o):=\mathcal{J}_F\pi(\hat{x}_i)^{\dagger} \mathcal{J}_F^{\dagger}=-\pi(\hat{x}_i^R)\label{N43}
	\end{equation}
	We  can therefore identify $\hat{x}_i^o=-\hat{x}_i^R$, rather than as $\hat{x}_i^o=\hat{x}_i^R$, as anticipated earlier. But the presence of the minus sign turns out to be a harmless, as it simply shows that elements of the opposite algebra  $\mathcal{A}_F^o$ is also now subjected to the parity transformation as
	\begin{equation}
		\hat{x}_i^o=-\hat{x}_i^R=P\hat{x}_i^R\label{N56}
	\end{equation} 
	This means that both $\hat{\vec{x}}^o$ and $\hat{\vec{x}}^R$ act from the right and thus belongs to $\mathcal{A}_F^o$, but while $\hat{\vec{x}}^R$ satisfies the commutator algebra (\ref{N55}) appropriate for right handed system, $\hat{\vec{x}}^o$ will satisfy that of left handed system. But since again the more general form of commutation relation is now given by (\ref{N55}), satisfying the entire O(3) symmetry, it immediately follows that $\hat{x}_i^o$ will also now satisfy the same SU(2)$^R$ Lie-algebra (\ref{eq60}):
	$$[\hat{x}_i^o,\hat{x}_j^o]=-i\lambda \epsilon_{ijk}\hat{x}_k^o$$
	It therefore follows trivially the following commutator relation is also satisfied.
	\begin{equation}
		[\mathcal{J}_F\pi(\hat{x}_i)^{\dagger} \mathcal{J}_F^{\dagger}, 	\mathcal{J}_F\pi(\hat{x}_j)^{\dagger} \mathcal{J}_F^{\dagger}] = [\pi(\hat{x}_i^R),\pi(\hat{x}_j^R)]= -i\lambda\epsilon_{ijk}\pi(\hat{x}_k^R) 
	\end{equation}
	Finally we would like to mention that our orbital angular momentum  (\ref{eq16}) flips sign under parity and is consistent with our new definition of angular momentum given in a commutative context ($\mathbb{R}^3$) as $L_i=E_{ijk}\hat{x}_j\hat{p}_k$ rather than as $\epsilon_{ijk}\hat{x}_j\hat{p}_k$. This forces the $\sigma_i$ occurring in the spin part to flip sign, thereby ,rendering the chirality  operator (\ref{eq8}) , an even object under parity. Further  if we replace $\epsilon_{ijk}$ by $E_{ijk}$ in the Dirac operator (\ref{eq8}) , it too becomes even under parity. 
	\subsection{Violation of first order condition}
	The requirement of first order condition given by $[[D_F,\pi(a)],\pi(b^o)]=0$ actually encodes the fact that the Dirac operator is a differential operator of order one and it is a derivation of the algebra $\mathcal{A}$, into itself, with the entire opposite algebra $\mathcal{A}^o=\mathcal{J}\mathcal{A}\mathcal{J^*}$ belonging to its commutant. In \cite{connes3}, it was shown that, by violating the first order condition, the fluctuated Dirac operator $\mathcal{D}_A$ becomes noninvariant under the inner fluctuation and to make it invariant again we have to add a qudratic inner fluctuation term to $\mathcal{D}_A$. \\
	To that note, let us check whether the condition is valid or not in our context. For that, let us choose, $\pi(a)=\pi(\hat{x}_l)=diag(\hat{x}_l,\hat{x}_l)$ and $\pi(b^o)=\pi(\hat{x}_p^o)=diag(\hat{x}_p^o,\hat{x}_p^o)$. Now using the form of the Dirac operator given in (\ref{eq8}), we can show,
	\begin{align}
		[[D_F,\pi(\hat{x}_l)],\pi(\hat{x}_p^o)]=\frac{i}{\lambda r_n}\gamma_F\Big[[ \epsilon_{ijk}\sigma_i\otimes\hat{x_j}^R\hat{x}_k,\textbf{I}\otimes\hat{x}_l],\textbf{I}\otimes\hat{x}_p^R\Big]&=-\frac{\gamma_F}{r_n}\Big[(\sigma_l\otimes \hat{x}_j^R\hat{x}_j-\sigma_i\otimes\hat{x}_l^R\hat{x}_i),\textbf{I}\otimes \hat{x}_p^R\Big]\nonumber\\
		&=-\frac{i\lambda\gamma_F}{r_n}\Big(\epsilon_{lpm}\sigma_i\hat{x}_i\hat{x}_m^R-\epsilon_{jpm}\sigma_l\hat{x}_j\hat{x}_m^R\Big)\ne 0
	\end{align}
	So the first order condition is violated. 
	\section{Discussion and future direction}
	In this paper we have tried to provide a consistent formulation of an even and real spectral triple for the fuzzy sphere in its 1/2 representation
	using Watamura’s prescription of Dirac and grading operator \cite{wat}. We started by obtaining the explicit expressions
	of the eigen spinors of SU(2) covariant forms of Dirac and chirality operators. Finally, we demonstrate how
	the real structure operator in spin-1/2 representation, consistent with the spectral data of KO dimension-4 can be obtained. This spectral triple for fuzzy sphere then should serve as the required mathematical structures necessary to build gauge theories on.\\ \\ In fact, since the above mentioned spectral triple violates the first order condition - an important ingredient in spectral formulation of standard model, this opens the door to investigate phenomena beyond standard model
	through a toy model where we can take the fuzzy sphere as the internal space along with 4D commutative manifold so
	that the algebra becomes $C^{\infty}(M,M_2(\mathbb{C}))$. This should help us to build SU(2) gauge theory using the formulation
	of almost commutative geometry \textit{a la’} Connes. Also, in this prescription, the Dirac operator of the fuzzy sphere, being SU(2) covariant, by default, can be regarded to be a fluctuated one to begin with. It
	thus remains to understand whether it will be a sensible 
	idea to fluctuate an already fluctuated Dirac operator. A
	deeper understanding of this and other related issues 
	are essential to build consistent models, which can perhaps  
	shed light on the physics beyond standard model. Besides, 
	the fully fluctuated Dirac operator should play a role in defining the “Higgs field”, with which one may be able to build some plausible cosmological model. For example, one may try to describe the interaction between gravity and dark matter candidates \cite{marcoli} at a higher energy scale.\\ \\
	%In \cite{paulo}, the authors has shown that, $S^2_*$ emerges dynamically from an SU(N) gauge theory on 4-dimensional manifold. We will infact take a bottom up approach in our case and will try to formulate non-abelian gauge theories by taking $n^{th}$-representation of the fuzzy sphere algebra as internal space.  
	Finally, it will be interesting to compare the effective action derived through spectral action principle  with that
	of the usual Kaluza-Klein theory, where spacetime manifold is taken in the product form $M_4 \times S^2$ \cite{salam}, where $S^2$
	is the commutative 2-sphere and to study their contrasting features.
	%In , authors had considered a 6 dimensional space - a product of 4 dimentional flat commutative space-time manifold with a 2 dimensional commutative sphere and have studied Einstein-Maxwell theory. Just as Kaluza Klein mechanism, the compactification occured due to dynamics. The theory although is non-renormalizable and deals with  unexpectedly high gauge coupling, in the same limit which is needed to discard the towers of massive modes occuring in the theory. It would be interesting to replace the space $S^2$ in the above theory with $S^2_*$ and study some contrasting features of the same.   
	\section*{Acknowledgement}
	The authors would like to thank, Prof. T. R. Govindarajan, for proposing this interesting problem to us. We would also like to acknowledge  Dr. Shane Farnsworth, Prof. Latham Boyle, Prof.  Fedele Lizzi and Prof. A. P. Balachandran for their useful comments and suggestions. A.C. would like to thank DST-India for providing financial support in the form of fellowship during the course of this work and P.N. would like to thank S N Bose National Centre for Basic Sciences, Kolkata, for providing financial support during the project tenure. Finally, A.C. and P.N. would like to thank Prof. Sibasish Ghosh for his generous support and hospitality in Intitute of Mathematical Sciences, Chennai, where this work was initiated.
	
\end{document}